\documentclass[twocolumn]{aastex7}
\usepackage{float}
\usepackage{graphicx}
\usepackage{amsmath}
\usepackage{natbib}
\usepackage{color}
\usepackage{verbatim}
\usepackage[utf8]{inputenc}
\usepackage{mathtools}
\usepackage{amssymb}
\usepackage{textcomp}
\usepackage{enumitem}
\usepackage{braket}
\usepackage{graphicx} 
\usepackage{tablefootnote}

\begin{document}

\title{Probing Turbulence, Gravity, Supernovae, and Magnetic Field Effects with the 6D Kinematics of Young Stars in Milky Way Star-Forming Regions} 

\author[0009-0001-1147-6851]{Benjamin N.\ Velguth}
\email{bvelguth@umass.edu}
\affiliation{Department of Astronomy, University of Massachusetts Amherst, Amherst, MA 01003, USA} 
\affiliation{Department of Physics, University of North Texas, Denton, TX 76203, USA} 

\author[0000-0001-5262-6150]{Yuan Li}
\email{yuanli@umass.edu}
\affiliation{Department of Astronomy, University of Massachusetts Amherst, Amherst, MA 01003, USA} 
\affiliation{Department of Physics, University of North Texas, Denton, TX 76203, USA} 

\author[0000-0001-6600-2517]{Trung Ha}
\email{tvha@umass.edu}
\affiliation{Department of Astronomy, University of Massachusetts Amherst, Amherst, MA 01003, USA} 
\affiliation{Department of Physics, University of North Texas, Denton, TX 76203, USA} 

\author[0000-0002-8455-0805]{Yue Hu}
\thanks{Hubble Fellow}
\email{yuehu@ias.edu}
\affiliation{Institute for Advanced Study, Princeton, NJ 08540, USA}

\author[0000-0002-5365-1267]{Marina Kounkel}
\email{marina.kounkel@unf.edu}
\affiliation{Department of Physics and Astronomy, University of North Florida, Jacksonville, FL 32224, USA}

\author[0000-0002-0458-7828]{Siyao Xu}
\email{xusiyao@ufl.edu}
\affiliation{Department of Physics, University of Florida, Gainesville, FL 32611, USA}

\author[0000-0002-6447-899X]{Robert Gutermuth}
\email{rgutermu@umass.edu}
\affiliation{Department of Astronomy, University of Massachusetts Amherst, Amherst, MA 01003, USA} 

\author[0000-0002-1253-2763]{Hui Li}
\email{lihui870222@gmail.com}
\affiliation{Department of Astronomy, Tsinghua University, Beijing 100084, China }

\author[0000-0003-2660-2889]{Martin D. Weinberg}
\email{mdw@umass.edu}
\affiliation{Department of Astronomy, University of Massachusetts Amherst, Amherst, MA 01003, USA} 

\begin{abstract}
The dynamics of star forming gas can be affected by many physical processes, such as turbulence, gravity, supernova explosions, and magnetic fields. In this paper, we investigate several nearby star forming regions (Orion, Upper Sco, Taurus, and Perseus) for kinematic imprints of these influences on the newly formed stars. Using Gaia DR3 astrometry and APOGEE DR17 radial velocities, we compute first-order velocity structure functions (VSFs) of young stars in galactic Cartesian coordinates in both 6D (3D positions and 3D velocities) and 4D (3D positions and each 1D velocity) to identify signatures of turbulence and anisotropic motion. We also construct 3D and 1D radial velocity profiles to identify coherent expansion trends, and compare stellar proper motions to plane-of-sky magnetic field orientations in Taurus and Perseus. We find that the VSFs are mildly anisotropic, with slightly different amplitudes, slopes, or features in different directions in several groups, but in general, they are all consistent with Larson's Relation at intermediate length scales, especially in less compact groups. In several cases, the VSFs exhibit features suggestive of local energy injection from supernovae. Radial velocity profiles reveal clear anisotropic expansion in multiple groups, with the most extreme cases corresponding to those with the most anisotropic VSFs. In Perseus, we find that the motions of young stars are preferentially perpendicular to the local magnetic field. We find multiple, overlapping causes in each group for the observed kinematics. Our findings support that young stars remember more than just the turbulent state of their natal clouds. 
\end{abstract}

\section{Introduction}
\label{sec:intro}

Star formation occurs in the densest part of the complicated and multiphase interstellar medium (ISM). Measuring the kinematics of the ISM is instrumental in our understanding of star formation process and galaxy evolution\citep[e.g.][]{Ballesteros-Paredes2020,Tacconi2020}. Much work has been done to observationally trace ISM motion in multiple phases, with most analyses suggesting that the ISM is broadly turbulent \citep{Larson1981,Lazarian2000,Heyer2004,Lazarian2006,ChepurnovnLazarian2010,Roman-Duval2011}.  

The main drawback of using only the kinematics of the gas is the limitation to line-of-sight velocities and non-negligible density fluctuations along the line of sight. The kinematics of the ISM are in three dimensions, and it is difficult to measure turbulence in the ISM without access to the six dimensional position and velocity information. Fortunately, inside some dense molecular clouds, star formation is taking place. It is expected that the newly-formed stars in these clouds retain the turbulent kinematics and clumpy distribution \citep{Kuhn2014,Sills2018} of their natal gas, making them potential ``tracer particles" of this gas. With the advent of the Gaia mission \citep{gaiaDR3} and access to more than just the position-position-velocity (PPV) information of gas, we can potentially probe the turbulent nature of star forming clouds by looking at the velocity statistics of young stars. 

\cite{Ha2021,Ha2022} (hereafter Ha21 and Ha22) were the first to put this idea into practice, using 3D positions and 2D proper motions from Gaia and radial velocities from APOGEE to look at the first order velocity structure functions (VSFs) of multiple nearby star forming regions. They found evidence for a Kolmogorov-like VSF at intermediate ($\ell \sim 10-100$ pc) length scales in the majority of the stellar groups they analyzed. This preliminary work is encouraging, but questions still remain regarding the origin of the power-law scaling they measured. 

Gravitational collapse/interactions, supernova explosions, tension and pressure from magnetic fields, and other forms of energy injection and pressure support from the local environment continuously alter the kinematics of the ISM and affect star formation \citep[e.g.][]{KoyamanInutsuka2000,KlessennHennebelle2010,Hennebelle2019,Hu2022}. \cite{Ma2025} find that in $\sim40\%$ of the molecular clouds they analyze, the VSFs of the gas deviate significantly from the expected ($v \propto \ell^{1/3 - 1/2}$) power law. They attribute this to the potentially non-negligible effect that local environments have on the turbulent cascade. Young stars, therefore, could also retain kinematic signatures of these effects. 

Recent studies of the kinematics of young stars have also highlighted anisotropic kinematics, particularly expansion trends, in young stellar associations \citep[e.g.][]{Armstrong2024,Sanchez-Sanjuan2024}. Anisotropy is not a characteristic of classical, hydrodynamical turbulence - observing anisotropy in the velocity statistics of these stellar groups would imply that the kinematics of stars reflect more than just the simple (Kolmogorov) turbulent cascade from the natal clouds. This requires us to further investigate the multi-scale velocity statistics of young stars in star forming groups. Do young stars ``remember" other dynamical influences on the ISM as well as the turbulent energy cascade? Are these effects contributing to the anisotropic kinematics that have been observed? 

In this work, we revisit the star forming regions analyzed in Ha21 and Ha22. We use a combination of Gaia DR3 and APOGEE DR17 survey data to measure anisotropic kinematics and expansion trends, probing the effects of ISM turbulence, cloud collapse, gravitational interactions, supernova explosions, and the local magnetic field on the kinematics of young stars. The paper is structured as follows: Section \ref{sec:datanmethods} outlines our kinematic data, stellar group assignments, and magnetic field maps. Section \ref{sec:methods} covers our methods to calculate the first-order VSFs, measure the expansion trends, and compare stellar proper motions with magnetic field orientations. Section \ref{sec:results} presents our analysis results of the VSFs, the expansion profiles, and the comparison between stellar kinematics and the magnetic field, as well as discussions of how each of these factors affects the kinematics of young stars. We describe the limitations of this study and future avenues of research in Section \ref{sec:final_limits_future}, and conclude in Section \ref{sec:conclusion}.

\section{Data Acquisition and Preparation}
\label{sec:datanmethods}

\subsection{Stellar Observations and Group Assignments}
\label{sec:obs}

We obtained the plane-of-sky positions, parallaxes, and proper motions of stars from the third data release (DR3) of the Gaia mission \citep{gaiaDR3} in the four nearby star forming regions: the Orion Molecular Cloud Complex, the upper Scorpius region (Upper Sco), Perseus (including NGC 1333 and IC 348), and Taurus. Line of sight velocities were observed with the Apache Point Observatory Galactic Evolution Experiment (APOGEE) spectrograph, mounted on the 2.5 meter Sloan Foundation Telescope of the Sloan Digital Sky Survey \citep{Gunn2006,Blanton2017}. The APOGEE-2 Data Release 17 collected spectral data and derived radial velocities for roughly 657,000 stars in the Milky Way, which we combined with the five-dimensional (5D) astrometry and proper motions from Gaia to obtain 6D (3D positions and 3D velocities) information for the stars used in this work. 

We identified stars in each region by performing a clustering analysis on the Gaia DR3 sources in the disk of the Milky Way using a Python implementation of HDBSCAN \citep[Hierarchical Density-Based Spatial Clustering of Applications with Noise;][]{hdbscan}. Binary stars were removed, and initial cuts in parallax (parallax $<$ 2), relative parallax error (parallax error/parallax $>$ 0.05), proper motions in $l$ and $b$ directions (proper motion $>$ 60 mas/yr), and galactic latitude (b $>$ 40 deg) were made to the larger sample of stars in DR3 \citep[following][]{Kounkel2019} before the clustering algorithm was run. We used the \texttt{leaf} method, a \texttt{min\_cluster\_size} of 180 sources, and a \texttt{min\_samples} value of 10. These choices were made to recover the large-scale coherent structures of stars within these star forming regions, at the expense of stars at the periphery of our groups. We split the Orion Molecular Cloud Complex into four sections: $\lambda$ Ori, ONC, Orion A, and the central region containing Orion B, C, and D  (hereafter Orion BCD). We analyzed each of these groups independently and as one larger group. We analyzed each of Taurus, Perseus, and Upper Sco as one group. We use ``groups" and ``clusters" in this work to refer to the associations of stars. We recognize that some of our groups contain sub-clusters, and some of the clusters analyzed here may not be fully bound.

\begin{figure*}[bht]
    \centering    
    \includegraphics[width=\linewidth]{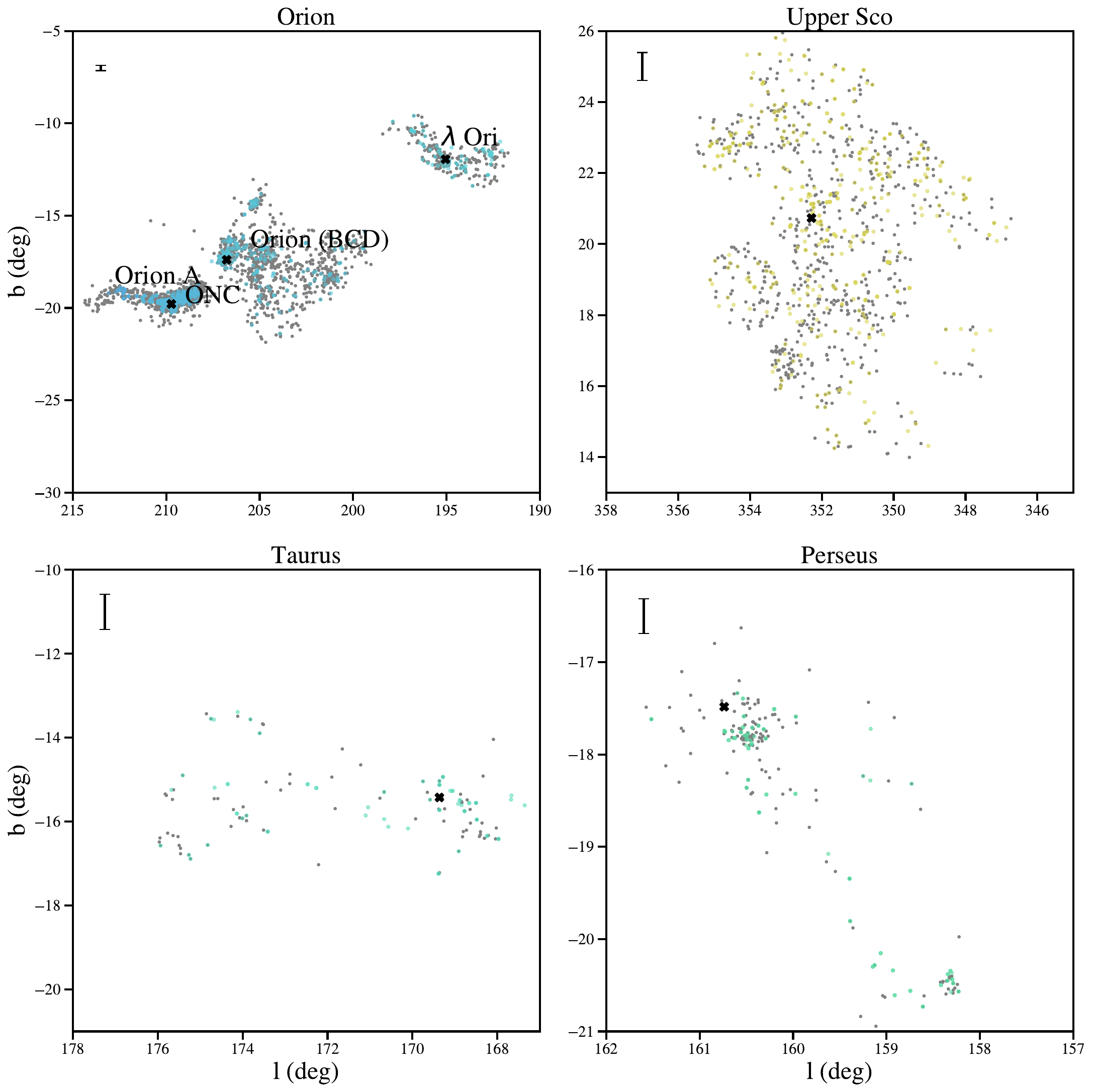}
\caption{The young stellar groups analyzed in this work in galactic coordinates. Colored points are the stars analyzed in this work, and gray points are those assigned to these groups and analyzed in \cite{Ha2021,Ha2022} (see Table~\ref{Tab:table} for sample comparison and Section~\ref{sec:datanmethods} for more details). The black x in each group is the location of the center of mass we estimate. Scale bars in the upper left of each panel indicate a physical distance of 1 parsec.}
    \label{fig:clusters_galactic}
\end{figure*}

We selected group members with membership probability above $25\%$ as defined by HDBSCAN. We then removed sources with radial velocity errors further from the mean than 1.5 times the mean absolute deviation (MAD), roughly 5 kms$^{-1}$ in each group (see Figure~\ref{fig:velo_err_dists} in the Appendix). We did not make cuts based on errors in proper motion; these measurements made by Gaia are much more precise (by about 1 order of magnitude). We also cut stars with a parallax outside of 2 times the MAD, ensuring that we were not including stars projected in front of or behind our groups.

We lastly removed stars in each group with a reported velocity greater than 5$\sigma$ away from the group mean velocity in any of the x, y, or z directions in the Local Standard of Rest converted to the galactic Cartesian frame. This set of cuts removed less than 10\% of the stars initially found to be in each group in the most extreme case. We report the standard deviation of the velocity in each direction in Table \ref{Tab:table}. We extensively tested many combinations of parameters fed into HDBSCAN and found that our results are insensitive to the exact choice of parameters. We also tested different secondary cuts based on error measurements, velocities, and parallaxes. Overall, we found that the results are ``noisier'' with a less strict cut, as one would expect, but the overall conclusions remain the same. We discuss limitations and uncertainties in detail in Section~\ref{sec:final_limits_future}.

With our grouping method and selection criteria, we recover most of the sources in Ha22 and expand the sample size each group by a factor of $\sim 1.5-3$ (see Table \ref{Tab:table}). Figure \ref{fig:clusters_galactic} shows the final selection of each group in Galactic coordinates. 

\begin{deluxetable*}{cccccccc}[bth]
\tablecaption{
    \textnormal{The number statistics of the groups we analyze in this paper compared with that presented in Ha22, as well as the velocity dispersion (standard deviation of the velocity) in each direction, the median stellar age, an estimate of the crossing time, and an estimate of the relaxation time of each group.}
    \label{Tab:table}
}
\tablecolumns{8}
\setlength{\extrarowheight}{4pt}
\tablewidth{\linewidth}
\tabletypesize{\small}
\tablehead{
\colhead{Group} & 
\colhead{Counts} &  
\colhead{Ha22} &
\colhead{$f$} &
\colhead{$\sigma_{vx,vy,vz}$ (kms$^{-1}$)} & 
\colhead{Median Age (Myr)} &
\colhead{$t_{cross}$ (Myr)} &
\colhead{$t_{relax}$ (Myr)}
}
\startdata
\hline
Orion (all) & 2895  & 2647 & 69\% & 5.21, 3.84, 3.46 & 2.6  & 9.8 & 652.4 \\ 
Upper Sco & 1350  & 742 & 89\% & 5.71, 1.45, 2.57 & 4.8 & 4.0 & 138.3 \\ 
Taurus & 192 &  85 & 84\% & 3.69, 1.70, 2.12 & 1.5 & 7.0 & 49.0 \\ 
Perseus & 374  & 182 & 67\% & 3.44, 4.97, 4.41 & 2.3  & 7.0 & 78.6 \\ 
\hline
\enddata
\tablecomments{The recovered fraction $f$ is the percentage of stars in Ha22 that are recovered in our group assignments by matching their on-sky positions within 1 arcsecond.
}

\end{deluxetable*}

\subsection{Magnetic Field Data}
\label{sec:Bdata}

To investigate the alignment between stellar proper motions and the projected magnetic field around these stars, we utilized maps of the POS magnetic field orientation derived from two independent methods: polarized dust thermal emission at 353 GHz observed by Planck \citep{2020A&A...641A...3P} and the Velocity Gradient Technique (VGT; \citealt{2019NatAs...3..776H}). The Planck observations provide the magnetic field angle $\phi$ through the Stokes parameter maps $Q$ and $U$: 
\begin{equation} 
\phi = \frac{1}{2} \arctan(-U, Q) + \frac{\pi}{2}, 
\end{equation} 
where the $-U$ term converts the angle from the HEALPix to the IAU convention, and the two-argument $\arctan$ function ensures the correct handling of angular periodicity. To enhance the signal-to-noise ratio, all maps were smoothed to a resolution of $10'$ using a Gaussian kernel.

The VGT is applied to the $^{12}$CO emission lines from the COMPLETE Survey \citep{2006AJ....131.2921R}. The VGT procedure consists of the following steps (see \citealt{2025ApJ...983...32H} for details): (1) selecting spectroscopic velocity channels that satisfy the "thin channel" criterion, where the channel width is smaller than the velocity dispersion \citep{2000ApJ...537..720L,2023MNRAS.524.2994H}; (2) convolving the thin channels with a Sobel kernel to produce raw gradient maps, blanking pixels where the intensity is less than three times the root-mean-square noise level; (3) calculating the local gradient orientation at each pixel using a sub-block averaging method \citep{2017ApJ...837L..24Y}, which statistically combines the orientations within a rectangular sub-block of the raw gradient map; and (4) constructing pseudo-Stokes $Q$ and $U$ parameters from the gradient orientations \citep{2020ApJ...888...96H}, enabling magnetic field orientation tracing in a manner analogous to the Planck observations. 

\begin{figure}[bht]
    \centering
    \includegraphics[width=\linewidth]{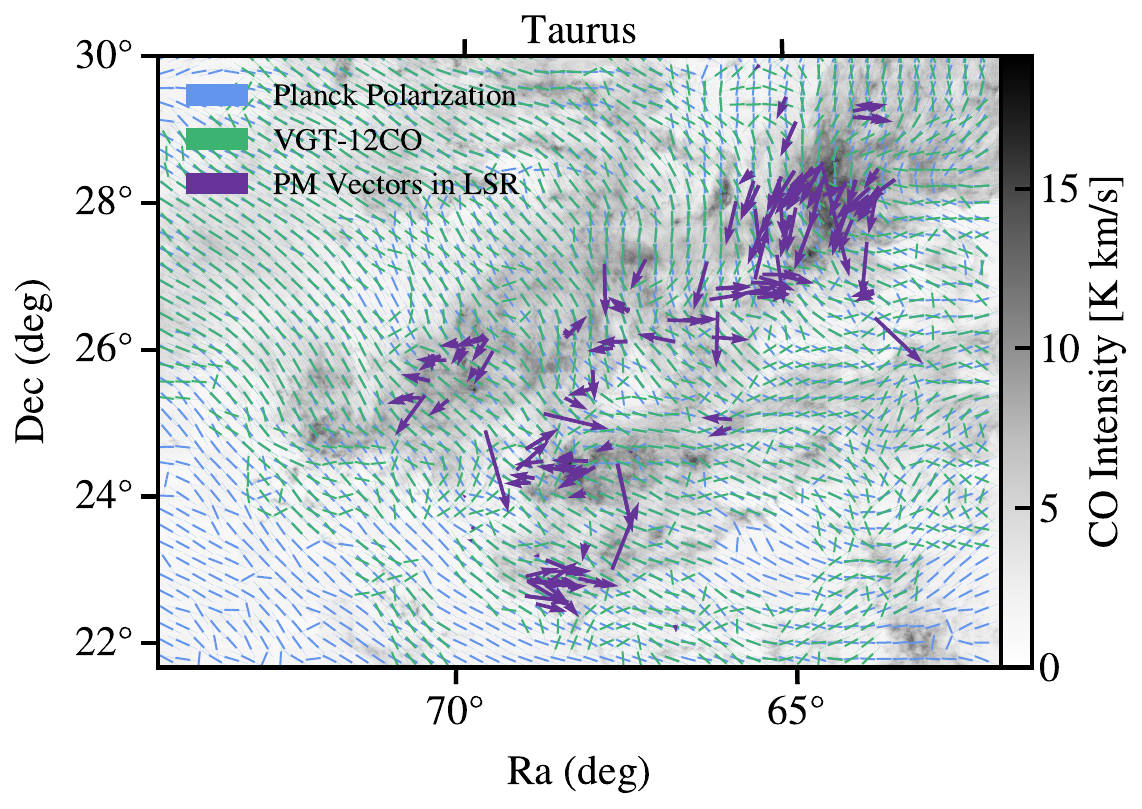}
    \includegraphics[width=\linewidth]{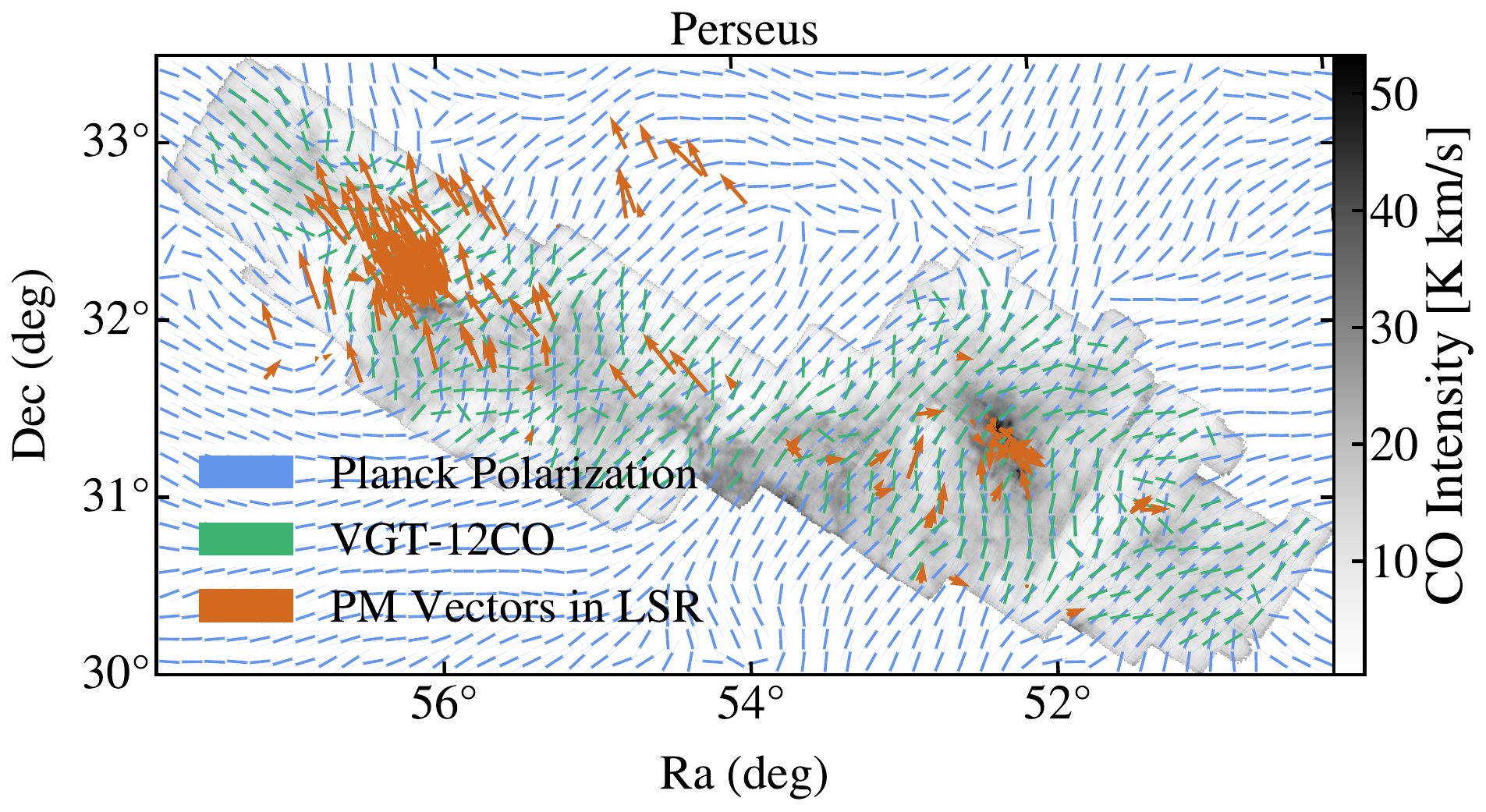}
    \caption{Taurus and Perseus star forming regions. Arrows show motions of young stars in the plane of the sky within these clouds. Gray background shows the CO intensity. Thin green and blue lines show the magnetic field orientations in the plane of the sky derived from VGT and Plank Polarization, respectively (see Section~\ref{sec:datanmethods} for more details).}
    \label{fig:B_field_maps}
\end{figure}

Maps of these orientations with the stellar proper motion vectors overlaid can be seen in Figure \ref{fig:B_field_maps}. We only look at the Perseus and Taurus groups here, as they are the only groups with the quality of magnetic field measurements necessary to perform this analysis. The resolution of VGT maps for Perseus and Taurus is $10'$ and $25'$, respectively. 

\section{Methodology}
\label{sec:methods}

\subsection{Velocity Structure Functions}
\label{sec:vsfs} 

The first order velocity structure function (VSF) is a diagnostic tool that is related to the kinetic energy power spectrum of a velocity field. It is a type of two point correlation function calculated as the mean absolute value of the velocity difference $\langle |\delta v| \rangle $ versus the physical separation $\ell$ between pairs of points. For a subsonic (incompressible) turbulent velocity field, we expect the slope of the power law scaling to be $1/3$ ($\langle |\delta v| \rangle  \propto \ell^{1/3}$) \citep{Kolmogorov1941}, and in highly supersonic and shock dominated (compressible) turbulence we expect a power law scaling of $1/2$ ($\langle |\delta v| \rangle  \propto \ell^{1/2}$) \citep{Burgers1995}. \cite{Larson1981} showed that Milky Way molecular clouds have a characteristic self-similar relation between their size and velocity dispersion ($\sigma \propto R^{0.38}$), suggesting that their motion is turbulent. A stellar group-wide expansion will manifest as a slope of roughly 1 ($\langle |\delta v| \rangle \propto \ell^{1}$). Higher velocity stars will drift further away from one another in a roughly linear trajectory and approach a Hubble-flow like expansion, with stars being born with higher velocities achieving this in roughly one crossing time ($t_{cross}$). Finally, a slope of zero ($\langle |\delta v| \rangle \sim $ constant) implies a dynamically relaxed group/cluster.

We computed the first order VSF in each region using both 3D position and 3D velocity data to obtain 6D VSFs. We also calculated these with 3D positions and one galactic velocity component to obtain 4D VSFs which we used to check for signatures of anisotropic kinematics. The 6D VSF is related to the 4D VSFs as $\delta v_{6D}^2 =(\delta v_x^2 + \delta v_y^2 + \delta v_z^2)$.

The VSFs were calculated as follows: First, the positions, distances, radial velocities, and proper motions of these stars were transformed into Cartesian, galactic coordinates with respect to the Galactic Local Standard of Rest (LSR), with x in the direction of galactic center, y in the direction of the galactic rotation, and z in the direction of the galactic north pole. Next the separations $d$ between every pair of stars were segmented into log-space bins $\ell$, and every pair within each bin was recorded with the absolute value of their velocity differences $|\delta v|$ calculated. This $|\delta v|$ was averaged to obtain $\langle |\delta v| \rangle$ for each bin of $\ell$. 

To compute the uncertainties in the structure functions, we preform random sampling of the measurements of each star. We obtain 100 realizations for each star following a Gaussian distribution, with the measured value being the mean and the reported error defining the standard deviation of the distribution. The radial velocity measurements from APOGEE are much more imprecise (by roughly an order of magnitude) and have a higher scatter than the proper motion measurements from Gaia (see Figure~\ref{fig:velo_err_dists}). To reduce systematic uncertainty, we first find the ratio of the width of the radial velocity error distribution to the proper motion error in right ascension and declination for each group. We then scale the proper motion errors of each star based on this ratio. Scaling the errors in this way preserves the relative precision of each star's measurements, while equalizing the noise level of the velocity measurements in each direction. We only consider uncertainties in parallax, proper motion, and radial velocity. Errors in right ascension and declination are much smaller than those of other quantities. With each iteration of the 100 VSFs, we excluded 10\% of the stars in each group from the calculation to take sampling statistics into account. We take the mean and standard deviation of each bin of $\ell$ as the group mean VSF and its corresponding error. 

We note that large bins of $\ell$ are poorly sampled, since these bins are approaching the size of the group. Drawing firm conclusions based on these under-sampled length scales is not advisable, and we only make inferences based on bins that are properly populated.

\subsection{Group Centers and Expansion Profiles }
\label{sec:exp_prof}

To define the center of each group, we adopted the iterative approach outlined in \cite{Armstrong2024} for three dimensional (as opposed to plane-of-sky) spatial data. First, a minimum inscribing sphere was generated around a given group, and the mean position of all group members in galactic Cartesian coordinates was determined. The sphere was then re-centered on this position and the radius of the sphere was shrunk by 5$\%$. The mean position of the stars remaining in the new sphere was found, and the next sphere was centered on this new mean position. This process was repeated until either the radius of the sphere fell below 0.5 pc or fewer than 3 stars remained inside the sphere. The locations of our final group center estimates are shown in Figure \ref{fig:clusters_galactic}.

We next calculated expansion profiles in each group from its center, with errors on each stellar parameter being calculated with the same resampling method outlined in Section \ref{sec:vsfs}. After subtracting the cluster median velocity vector from each star's 3D velocity vector, the radial components of each star's motion were calculated through taking the cosine of the angle between the unit position vector (relative to the group center) and velocity vectors of each star. We also computed expansion profiles along each direction (x, y, z).

We performed Markov Chain Monte Carlo (MCMC) fitting on each of the 3D and 1D expansion profiles with the Python package \texttt{emcee} \citep{emcee} with 200 walkers and 2000 iterations, with half of which being discarded as burn-in. We modeled linear fits in the same manner as \cite{Armstrong2022, Armstrong2024}, and assumed our errors to be Gaussian and independent. We selected wide, uniform priors for our fit parameters (\emph{m, b, f}) being slope, intercept, and fractional error of the linear relation. Errors in the location of each star were accounted for by varying the measured position according to its uncertainty during each iteration of the MCMC simulation as shown in \cite{Armstrong2024}. We compute the median, 16th, and 84th percentiles from the posterior distribution function as the linear best fit parameters and their respective uncertainties. 

\subsection{Magnetic Field Alignment}
\label{sec:Balign}
To quantify the alignment between magnetic fields and the stellar velocities, we measure the relative orientation of the magnetic field position angles with respect to the proper motion vectors of the stars. 

For each star, we match the on-sky position to the nearest magnetic field orientation, and then find the mean magnetic field angle within the 3x3 box that encloses the star to find the average field orientation in the vicinity of the star. We choose to average across a box enclosing the star to account for slight differences between adjacent field position angles and stars that do not fall directly on a grid position. These mean angles are then subtracted from the proper motion position angles and binned. We perform this calculation for both the Planck polarization and the VGT-derived maps described in Section \ref{sec:Bdata}. The mean Poisson error in each bin of $\delta\theta$ reflects the typical systematic uncertainty. Effectively, we are measuring the alignment between stellar motions and magnetic fields in the projected plane of the sky at a scale of $\sim3$ pc in both Taurus and Perseus.

\section{Results and Discussions}
\label{sec:results}

We explore three aspects of the stellar kinematics: in Section~\ref{sec:resTurb}, we study the connection between young stars and ISM turbulence; in Section~\ref{sec:resExpSN}, we compute the profiles of radial motions in young stellar groups to examine the effects of gravity (Section~\ref{sec:subsubGrav}) and supernovae (Section~\ref{sec:subsubSN}); we quantify the alignment between stellar motions and magnetic fields in Section~\ref{sec:resBfield}.

\subsection{Turbulence}
\label{sec:resTurb}

\begin{figure*}[bht]
    \centering    
    \includegraphics[width=\linewidth]{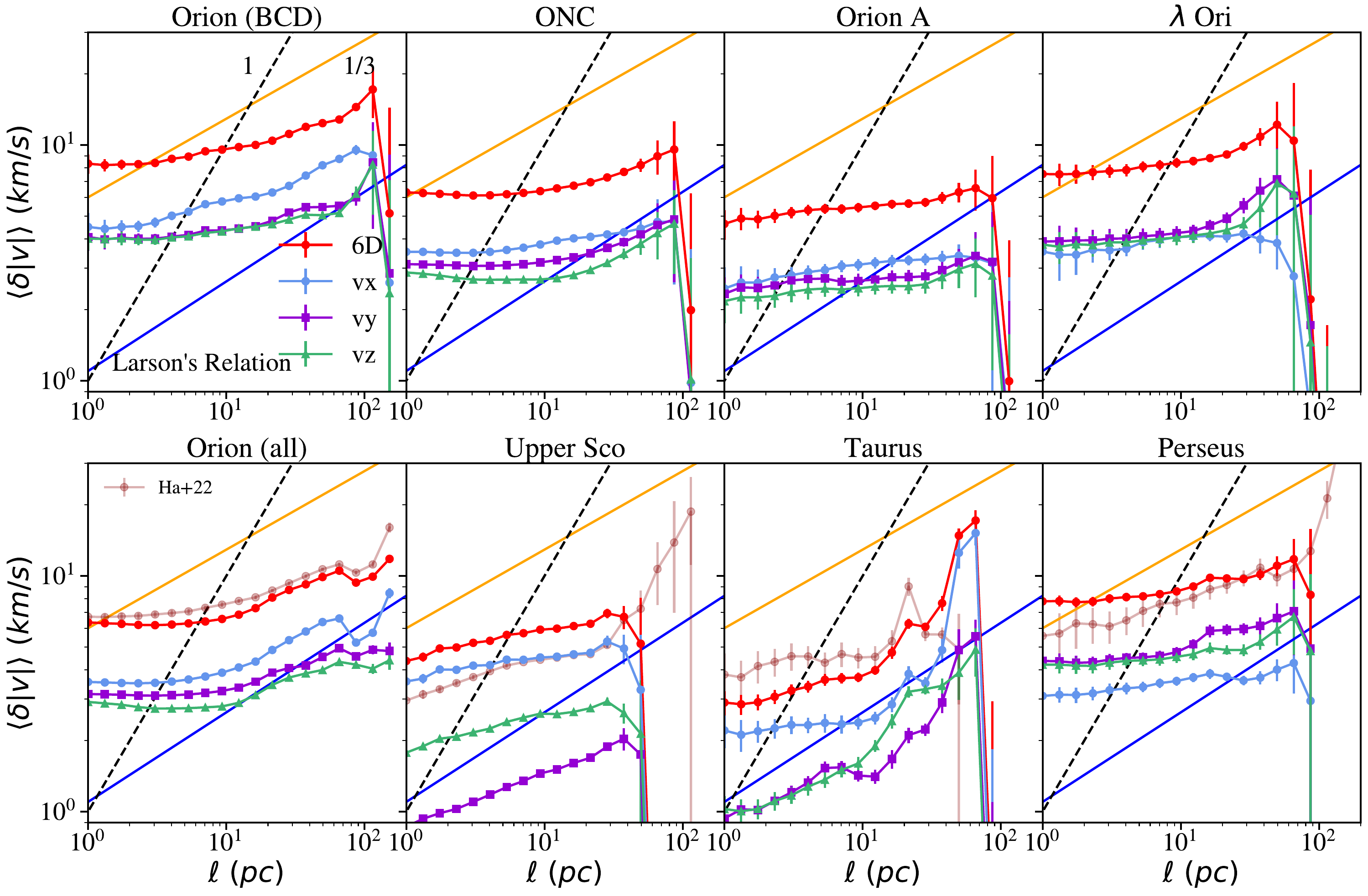}
    \caption{6D (red) and 4D (blue, green, purple) VSFs for each of the groups analyzed in this work. Also shown are the 1/3 (Kolmogorov, solid yellow line) and 1 (free expansion, dashed black line) power law slopes, along with Larson's Relation (solid blue line), for reference. }
    \label{fig:VSF}
\end{figure*}

To understand velocity statistics of the groups, we analyze the turbulent kinematics retained by these stars. We show both the full 6D VSFs and the three projected 4D VSFs in Figure \ref{fig:VSF}. Lines with slopes of $1/3$ and 1 for reference. We also show Larson's relation ($1.1\times\ell^{0.38}$) mainly as a reference for the amplitude of the ISM turbulence. Technically, there is a factor of order unity between VSF and $\sigma$ used in Larson's relation, where $\langle | \delta v | \rangle= \sqrt{\frac{2}{\pi}}\sigma$. We have omitted the factor for simplicity. In the bottom row, we show the 6D VSFs initially presented in Ha22. They select Ophiuchus out of Upper Sco in their work, but we include the entire region from their dataset to compare with ours. 

Overall, many of the groups show features consistent with local energy injection, such as bumps in the VSFs, and some of these VSFs show clear power law scaling at intermediate length scales, evidence of some level of retention of the turbulent kinematic state of their natal gas. All of the 6D VSFs show amplitudes generally consistent with Larson's Relation on large scales but higher on small ($\ell \lesssim 10$ pc) scales. 

Orion BCD's 6D VSF shows a clear power law scaling up on large scales, indicative of some retention of turbulence, albeit with a flatter slope at $\ell<20$ pc. ONC's VSF is very flat due to its dynamically evolved state. Orion A's slope in Ha21 was much steeper than we show here. We attribute this discrepancy to our sample selection difference. Gaia DR3 recovers more stars in Orion A, but many of them are spatially concentrated and are likely in small, bound clusters. For example, \cite{Sanchez-Sanjuan2024} recover two small, compact clusters in this region using DR3. Our sample also has a relatively strict cut in terms of membership probabilities and measurement/error cuts, which preferentially removes stars at the outskirts of the group (see Figure~\ref{fig:clusters_galactic}). As a result, our sample is more biased toward dense clusters rather than larger diffuse regions. We discuss our limitations and biases further in Section~\ref{sec:final_limits_future}.

At small scales, the 6D VSF in $\lambda$ Ori shows a flat slope at small $\ell$ that transitions into a power law of $\sim 1/3$ over a short dynamical range (from $\sim20$ to $\sim 60$ pc). The VSF then peaks at $\ell \sim 60$pc, likely due to the expansion that has been previously observed in this cluster's outskirts \citep{Armstrong2024}. See Section~\ref{sec:resExpSN} for further discussions of this trend. 

The 6D VSF of the entire Orion group shows each of the above trends in superposition. The flat trend at small $\ell$ is likely due to the dynamical relaxation of stars in bound clusters (dominated by ONC here). The power law scaling at intermediate $\ell$ reflects the turbulent nature of these stars' natal gas, and the peak at higher scales is evidence of expansion, from large-scale dynamical effects and/or a recent SN explosion. 

The 6D VSF of Upper Sco is relatively flat up to $\ell \sim 30$ pc, where it peaks and then turns over. This peak is possible kinematic evidence for a past supernova contained in the Ophiuchus region \citep{Neuhauser2020}. Taurus's VSF shows a steeper slope at shorter $\ell$ and a small peak at $\sim$25 pc. This is possibly indicative of local energy injection likely from a SN explosion (see Section~\ref{sec:resExpSN} for a discussion of these feature).

We note that the crossing time we estimate in Upper Sco is roughly that of the median stellar age. In contrast to Ha22, who display a steepening at large $\ell$, we see a flat slope at smaller scales (due to subclusters in the larger group) and do not sample larger bins of $\ell$ unless we are much less restrictive with our parallax cut. It is possible that this difference is due to higher-velocity stars drifting to higher radii within a crossing time, which we would remove with a strict cut, but could also be caused by contaminates along the line of sight being included in the Ha22 catalog. Forthcoming work will bin each of our groups by age to determine the time evolution of these stellar kinematics to potentially recover the Komolgorov-like slope in the youngest populations in Upper Sco, as well as trace back stellar trajectories over time to help determine which stars truly originate from this star forming region.

Perseus's 6D VSF shows weak power law scaling at all length scales, with two distinct peaks at $\sim$20 and 80 pc. The peak at $\sim$20 likely reflects the relative motion between the two sub-clusters within Perseus that are separated by $\sim 20$ pc (with a projected separation of $\sim 7$ pc in Figure~\ref{fig:clusters_galactic}). The peak at $\sim 80$ pc may be caused by a SN (see Section~\ref{sec:resExpSN} for further discussions). The relatively flat 6D VSF on small scales and its elevated amplitude at small $\ell$ compared with \citet{Ha2021} are likely caused by the sample selection difference, similar to Orion A as discussed previously.

To investigate the (an)isotropy of stellar kinematics, we also compute the 4D VSFs in x, y, and z directions, shown in blue, purple, and green, respectively, in Figure \ref{fig:VSF}. While some groups show generally consistent VSFs in different directions (e.g., Orion A), some groups show drastically different VSFs along different directions, such as $\lambda$ Ori and Taurus. The differences can be in the amplitudes, the slopes (e.g. Upper Sco) and/or the features (e.g. $\lambda$ Ori, Taurus) of the VSFs. 

The difference in amplitudes of these VSFs can be caused by the different velocity measurement uncertainties of the stars, instead of actual anisotropies. The larger uncertainty in the line of sight velocity measurements will produce a wider dispersion in one direction, which in turn will produce a heightened 4D VSF amplitude in that direction (even for a truly isotropic velocity distribution). In all of groups we analyze, the line of sight roughly translates to the x-direction in the coordinate system we adopt. The 4D VSF and velocity dispersion in this direction are equal to or higher than the 4D VSFs (Figure~\ref{fig:VSF}) and velocity dispersions in the other directions (Table~\ref{Tab:table}) in all groups but Perseus.

However, the different slopes and features in different directions in several groups are obvious (e.g., $\lambda$ Ori and Taurus). This is clear evidence for anisotropic motions that would not be caused by observational effects. 

The anisotropy (along the x, y, and z directions) in the 4D VSFs means that we are likely not observing exclusively pure turbulent kinematics. The broad agreement between our measured VSFs and Larson's relation suggest that turbulence is present in young stars' kinematics \citep[e.g.][]{Larson1981,ChepurnovnLazarian2010}. However, on top of this turbulence, supernovae can provide an additional (potentially anisotropic) kick to the star forming gas (changing the initial velocity). Moreover, gravity can also alter the kinematics of both star-forming gas and young stars on various timescales. In addition, the very steep VSFs on larger scales in $\lambda$ Ori (y and z directions) and Taurus are indicative of a group expansion, which is bulk motion rather than turbulence. It is possible that some of the groups have their kinematics dominated by large-scale expansion (corresponding to slope of 1 in the VSF) and small scale flattening from dynamical relaxation, which in combination show a Kolmogorov-like slope. We analyze the radial motion of each group in more detail in the following section.

\subsection{Radial Motion: Gravity and Supernovae}
\label{sec:resExpSN}

\begin{figure*}[bht]
    \centering    
    \includegraphics[width=\linewidth]{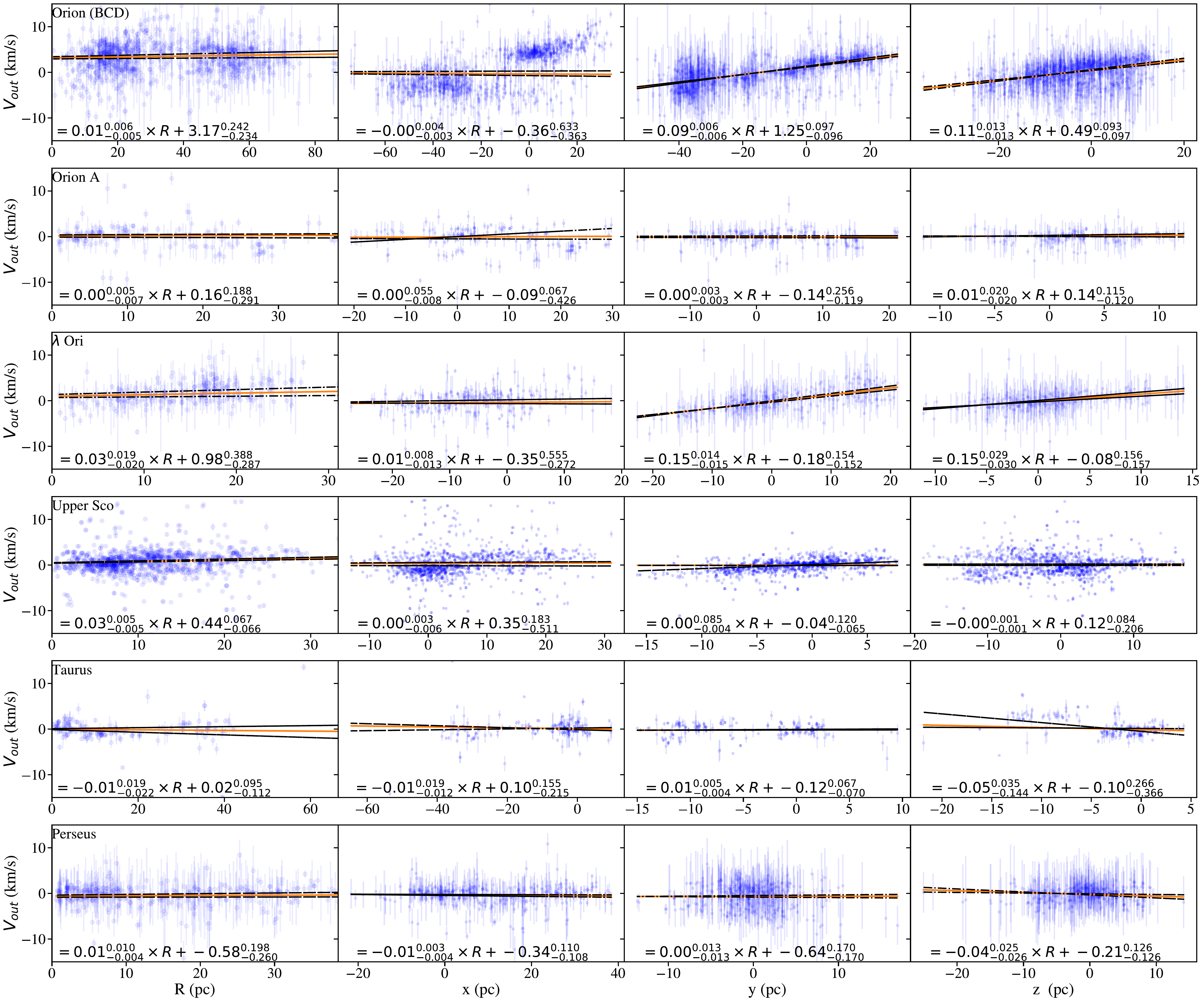}
    \caption{Expansion profiles of each of the groups analyzed in this work, with the 3D profile shown in the left column and each of the 1D profiles in the next three columns. Best fit lines are shown in orange, with 16th and 84th percentiles shown in the black dashed lines. The expressions including uncertainties are included at the bottom of each panel.}
    \label{fig:exp_all}
\end{figure*}
In addition to the turbulent cascade, gravity and SN explosions also directly affect the kinematics of the ISM and hence newborn stars. To explore the effects of gravity and SN explosions, we study the 3D and 1D expansion profiles of the stellar groups, shown in Figure \ref{fig:exp_all}. The different groups show a variety of different profiles, and many groups show diversity in their profiles along different directions. We do not show the profiles for the ONC and Orion, as the ONC is a bound, relaxed cluster and Orion is the combination of the other subgroups shown in rows 1-3 (and the ONC). 

The expansion profile of Orion BCD (top row of Figure~\ref{fig:exp_all}) shows two distinct groups that are both expanding away from the group center (most clearly seen in the x direction). This group is composed of multiple subgroups that have been previously shown to be expanding away from one another \citep{Kounkel2018,Sanchez-Sanjuan2024}. \cite{Kounkel2020} suggest that the expansion of these subgroups is driven by a SN explosion occurring roughly 6 Myr ago. Orion A, on the other hand, shows rather flat profiles in all directions, consistent with no clear expansion or contraction.

We find $\lambda$ Ori to be expanding, and the expansion is predominately in the y and z directions. This is qualitatively consistent with \citet{Armstrong2024}, who analyzed the expansion profile in proper motion space only. Our 3D expansion profile shows a shallower slope than their 2D profile, likely because the line-of-sight direction is mostly aligned with x and does not show clear expansion. 

Upper Sco appears to be mildly expanding in the 3D profile, with no prominent expansion appearing in the 1D profiles. It is possible that older stars that have migrated further out of this region were removed by our clustering algorithm choices and strict cut in parallax as mentioned previously. 

Taurus's profiles show evidence for contraction in the 3D profile and the x and z directions, with a slight expansion in the y direction. This behavior is unique among the groups analyzed in this work. Perseus shows a slight expansion in the 3D profile and a variety of profiles in each direction, but a high velocity scatter in each direction. The slope we see in the 3D profile has high enough uncertainty that we do not believe it to be a true, group wide expansion, especially in light of the flat VSFs and high velocity scatter in each direction. 

Many studies \citep[e.g.][]{Wright2019,Armstrong2022,Armstrong2024,Sanchez-Sanjuan2024} find strong evidence for anisotropic expansion, as we do, in young star forming regions, but the origin of this expansion is unclear. In the past, the most commonly cited mechanism for the disruption and expansion of a young cluster is gas expulsion: after a short episode of star formation, feedback from the young stars expels the remaining gas, halts star formation, and leaves the stars in the system in a supervirial state \citep{Lada1984}. However, this simplistic model of group dissipation predicts a spherically symmetric expansion of the stars after a single star formation episode, which is inconsistent with the presence of remaining gas, highly directional expansion, and more complicated star formation history these groups posses. We investigate two potential causes of these expansion trends: multi-scale gravitational effects (Section~\ref{sec:subsubGrav}) and the complex interactions between supernovae, the star forming environment, and multiple generations of star formation (Section~\ref{sec:subsubSN}).

\subsubsection{Gravitational Collapse and Other Dynamical Effects}
\label{sec:subsubGrav}

The formation of stellar associations in cool, overdense regions of the ISM is inherently fractal, where larger clouds fragment into smaller and smaller clumps down to the scale of individual stars \citep{LadanLada2003}. Gravity can also amplify anisotropies in the spacial distribution of matter, as gravity is stronger in denser regions and collapse happens preferentially along the shortest axis of a clump. The primary proposed formation mechanism for the filamentary structures we observe in the ISM is a hierarchical collapse of a cloud first into a sheet and then into a filament \citep{Hennebelle2019}. These filaments play an important role in the formation of stars, and dominate the mass budget of molecular clouds at high densities \citep[][and the references therein]{Ballesteros-Paredes2020}. They also have been observed to be enveloped in local magnetic fields \citep[][see Section \ref{sec:resBfield} for further discussions of the relationship between these magnetic fields and stellar kinematics]{StutznGould2016}. This collapse along different axes at different times can be imprinted into the kinematics of the stars they form.

As the collapse proceeds, stars form and decouple from the cloud. Later, individual stars can cross the group center and their radial (from the group center) velocities become positive such that the group appears as though it is expanding. In Upper Sco, the crossing time we estimate is roughly that of the median stellar age (Table~\ref{Tab:table}), implying that the slight expansion is at least partially due to the stars passing the group center. We see moderate expansion in the 3D profile and mild, but uniform, 1D expansion profiles (Figure~\ref{fig:exp_all}), implying a somewhat symmetric expansion of this group. In contrast, Taurus shows steepening in the 4D VSFs to $\sim1$ at varying length scales, and displays different expansion profiles in different directions. 

Taurus's clouds are quite massive \citep[$\sim 2.4 \times 10^4 M_{\odot},$][]{Goldsmith2008}, but stars have not been forming here for very long and there are no OB stars in this region \citep{Kenyon1995,Soler2023}. The directional inconsistencies in the 4D VSFs and 1D expansion profiles imply a memory of the hierarchical natal cloud collapse. Based on this, our findings are consistent with the scenario that clouds in Taurus are still assembling, and this group is in an earlier evolutionary stage than the others analyzed in this work. 

Within the larger groups, subgroups birthed in the same natal cloud can also form independently and move within the larger association. Orion BCD is composed of three subgroups, two of which (Orion C and D) are moving apart from each other along the line of sight (the x direction). The motion of these two subgroups is clearly displayed in the expansion profiles shown in Figure \ref{fig:exp_all} and likely contributes to the heightened VSF in the x-direction (Figure \ref{fig:VSF}).

The Perseus group is also composed of two sub clusters (NGC 1333 and IC 348, seen in Figure \ref{fig:clusters_galactic}). \cite{Kounkel2022} find that these two clusters are associated with larger molecular gas structures passing each other by. We attribute this peak to the relative motion of the two clusters in this group, as the peak appears at the 3D separation between these clusters ($\ell \sim 20$ pc). \cite{Kounkel2022} also find evidence for a SN explosion occurring in this region 1-2 Myr ago. We discuss this further in Section \ref{sec:subsubSN}.

Subgroups and sub clusters can also collide within the larger group, causing changes in the dynamical states of the stars. It is believed that many massive star clusters form out of this hierarchical assembly \citep{Cournoyer-Cloutier2024} as opposed to the hierarchical fragmentation that builds smaller sub-groups in single molecular clouds \citep{LadanLada2003}. \cite{Cournoyer-Cloutier2024} show that in the event of a sub-cluster merger event, stars that are ejected as a result of the merger show preferential directions of motion. \cite{Armstrong2024} conclude that the anisotropic expansion profile they see in $\lambda$ Ori is likely to be (at least) partially a result of a series of sub group mergers. Our expansion profiles agree with \citet{Armstrong2024} in projection. However, it has been hypothesized \citep{Mathieu2008,Kounkel2018} that $\lambda$ Ori's expansion is driven by an recent supernova, and we see a peak in the VSFs at $\sim$60 pc indicative of local energy injection. Further analysis of the kinematics of the subgroups contained in each larger region may uncover which mechanism is the dominant driver of the expansion.  

The gravitational field of the Galaxy can also influence stellar kinematics. While the time scale of epicyclic motions is long, the period of vertical oscillations in the solar neighborhood, $P_v\sim 90$ Myr, is shorter. The star forming regions in this study are all near the mid-plane of the Milky Way. The fractional change in the vertical velocity scales as $cos(2\pi t/P_v)$. For the majority of our groups with median stellar ages around 2 Myr (see Table~\ref{Tab:table}), the velocity change is only $\sim 1$\%. For Upper Sco with a median age of 4.8 Myr, the change is $\sim 6$\%. Thus, Galactic dynamics does not have a significant impact on our current results, but future studies of older populations should account for this effect, especially when interpreting expansion anisotropies.

\subsubsection{Supernova Explosions}
\label{sec:subsubSN}

Supernova explosions (SN) are commonly cited as one of the primary drivers of turbulence in the ISM \citep[e.g,][]{Joung2009,Padoan2016,Gent2020,Chamandy2020}. These events inject a massive amount of energy and momentum into the surrounding gas, creating bubbles of hot gas and ``snowplowed" shells of denser gas along the shock front that then can form a subsequent generation of stars. Stars that form on these shells inherit the expansion of the shells in their kinematics. These expansions manifest in our VSFs as bumps in a relatively narrow range of length scales $(\ell)$. The $\ell$ at the peak of this bump and its corresponding $\langle| \delta v | \rangle$ can be used to estimate the age of the SN ($\ell / \langle| \delta v | \rangle$). 

Ha21 find evidence for SN as a peak in the VSF of Orion ($\ell \sim 70$ pc). \citet{Kounkel2020} estimate the age of this SN to be roughly 6 Myr. We observe a similar bump in Orion at the same length scale and based on this $\ell$ and its corresponding peak $(\langle|\delta v|\rangle \sim 10 \ kms^{-1})$, we estimate the time since this SN to be roughly 7 Myr. We see evidence of this local injection of energy in each 4D VSF as well, implying that stars formed on this shell are roughly spherically distributed on its surface. This is reinforced by a positive slope in the y and z expansion profiles in Orion BCD (Figure~\ref{fig:exp_all}) and two clumps (Orion C and D) in the x direction moving away from one another. Taking the inverse (with a factor of 1.023 for unit conversion) of the steepest expansion slope, we obtain and upper limit of the time since this expansion began to be $\sim9$ Myr. The consistency of these timescales leads us to believe that the expansion we see here is driven by stars that formed on the expanding shell of a past supernova. 

In $\lambda$ Ori, we see a peak in the VSFs at $\sim$60 pc. Based on this we estimate the age of this SN to be $\sim$6 Myr, similar to age estimates given in \cite{Mathieu2008,Kounkel2018} for the same event. Furthermore, converting the best fit slope of the steepest expansion profile (z-direction) gives an upper limit of the time since the expansion began of $\sim6.8$ Myr. This is consistent with the timescale from our VSF and the upper limit of the expansion timescale given in \cite{Armstrong2024} (5.637 $\pm \sim$1 Myr). It is worth noting that the timescales of the expansion given in \citet{Mathieu2008,Kounkel2018} and the upper limits given in \citet{Armstrong2024} are consistent with one another, but have different proposed mechanisms (past SN versus subgroup collisions/mergers) causing the expansion. Future work is needed to determine when each of these processes occurred with higher accuracy, as well as uncover which process is the dominant driver of the expansion. 

In Upper Sco, we see a clear turnover in the VSFs at $\ell \sim 30$pc, which gives a timescale of $\sim4$Myr. Previous work \citep{Neuhauser2020} find that in the Ophiuchus subgroup, contained in Upper Sco, there was a supernova explosion roughly 2 Myr ago. Ha22 also find a peak in the VSF of stars younger than 2 Myr ($\ell \sim 25$pc) which they attribute to the same event. The timescale we derive is different than those found previously, and we do not see a strong expansion trend for this group. However, our inclusion of stars of all ages in our analysis has contaminated this peak, and it is possible that expansion trends driven by this SN are buried by the older ($>4$Myr old) population. This peak is also at the largest $\ell$ we probe, meaning these bins are poorly sampled. Future work with Gaia DR3 to reproduce this age-binned analysis in Ha22 will help resolve this discrepancy in derived timescales. 

\cite{Kounkel2022} find evidence for a past SN explosion occurring in the Perseus region, likely originating from one massive star in a binary system exploding less than 1-2 Myr ago. We see a peak in the Perseus VSFs at roughly 60 pc, and find a corresponding peak $\langle| \delta v | \rangle$ of $\sim10-15$km/s, giving a rough age estimate of 4-6 Myr. This estimate is inconsistent with the timescales found by \cite{Kounkel2022}. However, \cite{Kounkel2022} find no evidence of this SN triggering any star formation, which would explain why we do not see it imprinted on the kinematics of the stars. The peak we see could instead be an artifact, due to this $\ell$ approaching the size of the group which limits our ability to effectively sample this range of physical separations. 

\begin{figure}[bht]
    \centering   
    \includegraphics[width=\linewidth]{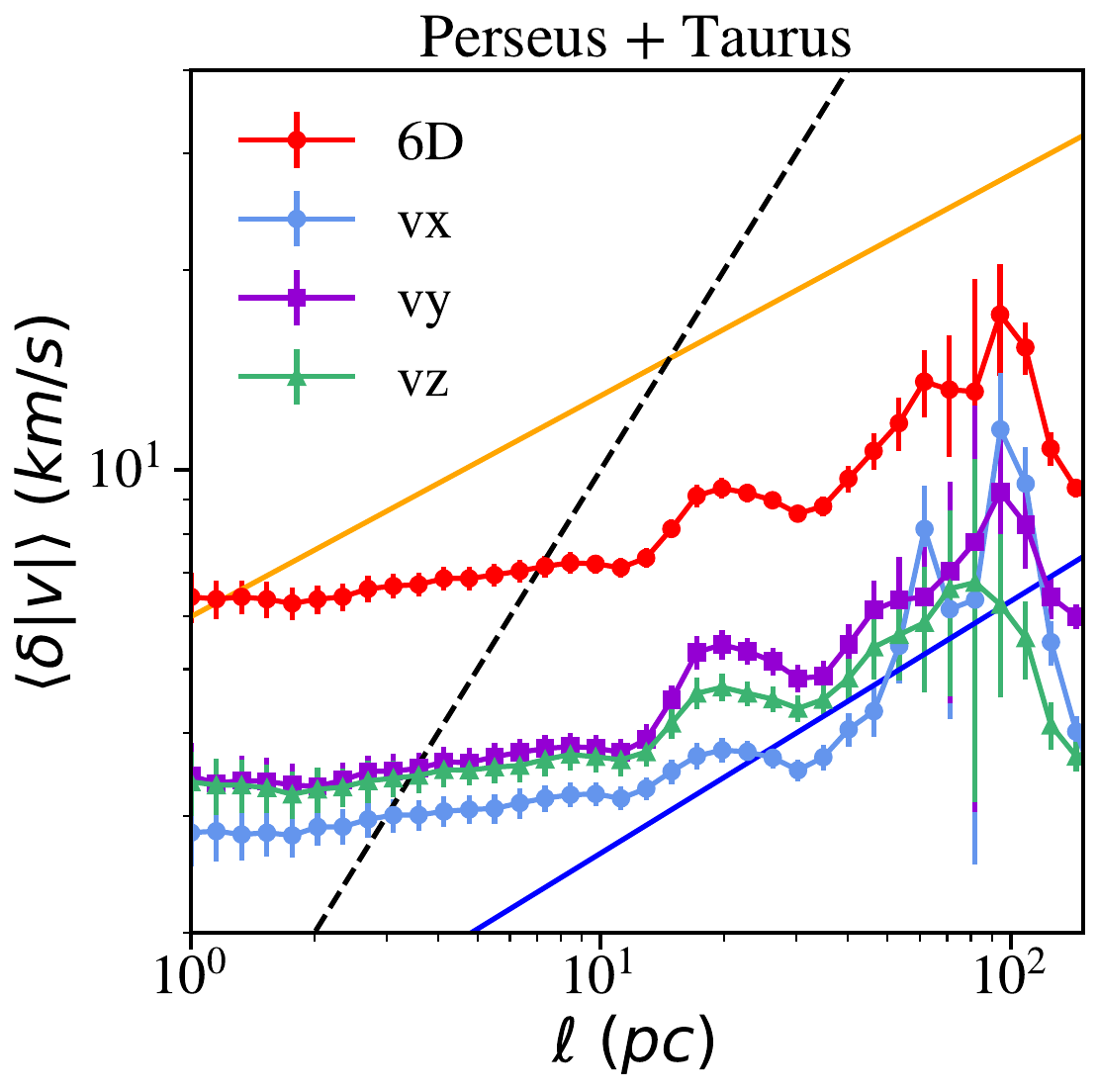}
    \caption{6D (red) and 4D (blue, green, purple) VSFs for Perseus and Taurus combined. The same reference lines are plotted here as in Figure~\ref{fig:VSF}. The peaks at $\ell \sim 20$ and $\ell \sim 100$ pc are indicative of energy injection from supernovae.}
    \label{fig:Per_taur_vsfs}
\end{figure}

\cite{Bialy2021} present evidence that the Perseus and Taurus molecular clouds formed on opposite sides of an extended shell (the ``Per-Tau shell", R $\sim$ 160 pc) driven by previous stellar and SN feedback. To see if we recover evidence of this event in the stellar kinematics, we create a 6D VSF of stars combining both Taurus and Perseus, seen in Figure \ref{fig:Per_taur_vsfs}. There is a prominent peak at $\ell \sim$ 100 pc with a corresponding $\langle | \delta v | \rangle$ of $\sim 20$ km/s, providing an age estimate of $\sim5$ Myr which is consistent with the lower limit of the age presented in \citet{Bialy2021} ($\sim$6-22 Myr). The wider uncertainties and slight dip in the VSF at the peak is due to the separation between these groups, as we cannot sample this length scale effectively since there are few members separated by this distance, even after our resampling of parallaxes based on the reported errors described in Section \ref{sec:datanmethods}. 

\cite{Bialy2021} also find a smaller shell within the Taurus cloud (the ``Tau Ring"), attributed to a second SN from a subsequent generation of stars formed on the larger, older shell. Furthermore, \cite{Krolikowski2021} mention that a SN could be the source of energy injection that they observe in their inter-core-group VSF at a $\ell$ of roughly 15 pc. We see evidence of local energy injection in our Perseus and Taurus combined VSF, as well as the VSF of only Taurus, both located at $\ell \sim 20$pc which matches this length scale as well as the radius of the Tau Ring reported in \cite{Bialy2021} (semimajor axis of 39 pc, semiminor axis of 26 pc). Taking our $\ell$ and its associated peak $\langle|\delta v|\rangle$, we estimate the age of this SN to be $\sim 2-3$ Myr. This is consistent with the idea of this SN being from a second generation of star formation caused by the pile up of molecular gas and subsequent generation of star formation along the first SN's shock front. 

\subsection{Magnetic Field Alignment}
\label{sec:resBfield}

Magnetic fields are ubiquitous in the ISM \citep[e.g.][]{Han1994,Han2009,Dickey2022}. These fields can affect the star formation process in tandem with gravity and turbulence \citep{McKee2007,Li2021}, as they provide both additional support against collapse, are warped and tangled by the bulk and turbulent motions of dense clouds, and suppress non-turbulent gas motion perpendicular to their alignment \citep[][and the references therein]{Hennebelle2019}. Here we investigate how aligned the local magnetic fields are to the proper motion vectors of the stars in our groups in the plane of the sky at $\sim 3$ pc scales. The lack of line-of-sight magnetic field direction prevents us from comparing with the 3D velocity of our stellar sample. 

\begin{figure}[bht]
    \centering    
    \includegraphics[width=\linewidth]{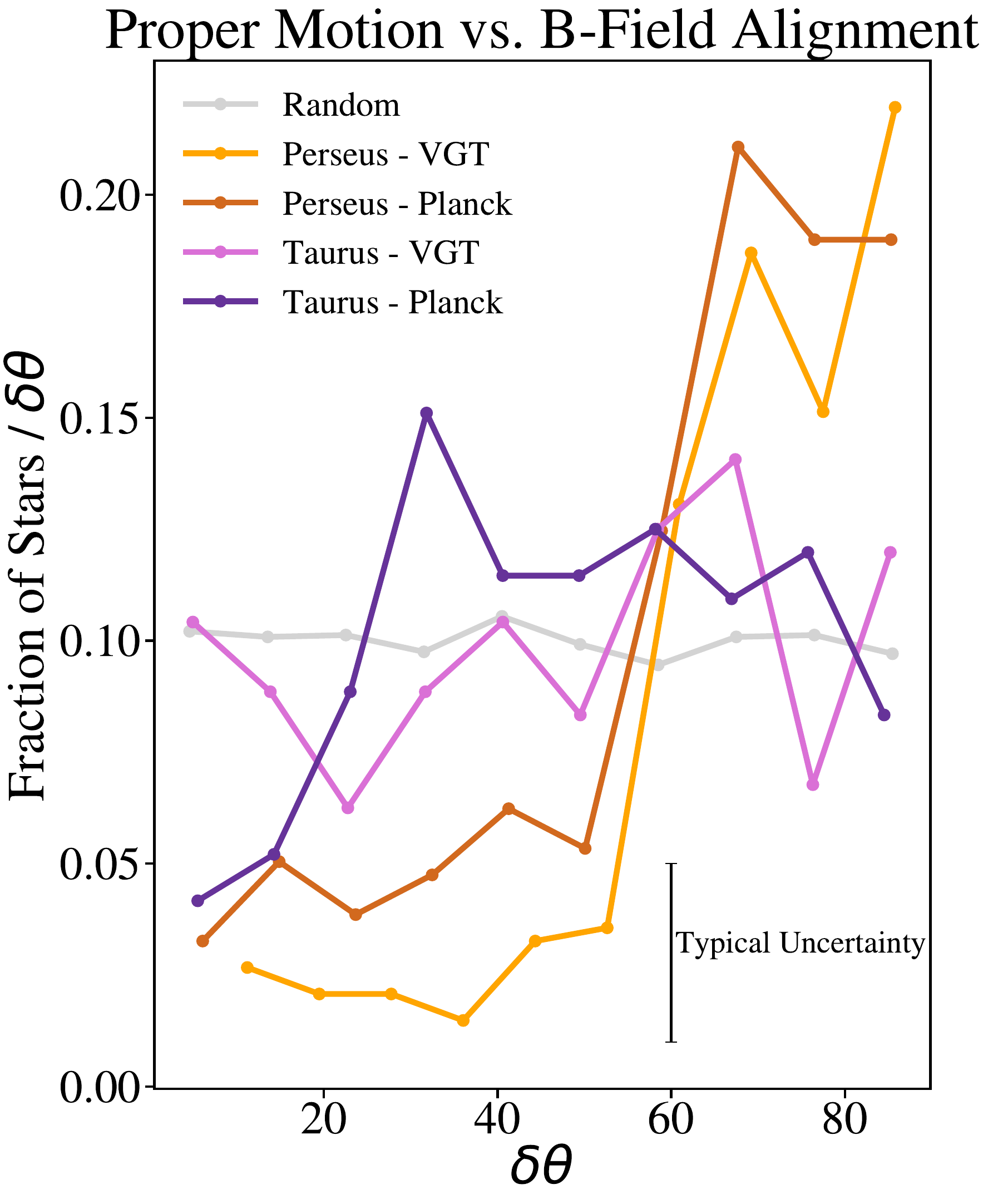}
    \caption{The distribution of relative angles ($\delta\theta$) between the proper motion vectors of young stars and the magnetic field orientations in their immediate vicinity in the plane of the sky in the Taurus and Perseus groups. We use magnetic fields measured with both VGT and Planck Polarizations in each region (see Section~\ref{sec:datanmethods} for details). Stellar motions are preferentially perpendicular to the magnetic fields in Perseus.}
    \label{fig:B_field_pdfs}
\end{figure}

Figure \ref{fig:B_field_pdfs} shows the alignments of magnetic field orientations and stellar velocities for both the Taurus and Perseus groups. We analyzed magnetic field measurements from two sources (outlined in Section~\ref{sec:Bdata}). Typical uncertainties $(\pm2\%)$ are representative of the systematic uncertainties. We calculate these by finding the mean Poisson error in each bin of $\delta\theta$ across both groups. To test if the trends we see are random, we generate 10000 position angles, randomly place them on the magnetic field map, and do the same calculations outlined in Section~\ref{sec:Balign}. We only show the results of this test for one map, but testing against both field measurement maps in both groups show the same result. 

We observe two distinct trends in these groups: in Taurus, the proper motion vectors appear to be randomly oriented relative to the magnetic field lines, and the stars in Perseus appear to be moving preferentially perpendicular to the field lines. Using the two sample ks-test, we find that the observed distributions of $\delta\theta$ in Taurus are not inconsistent with the random distribution (p $\sim$ 0.1) and the distributions in Perseus are inconsistent with the random distribution (p $\sim$ 0.01).

Many works \citep[e.g.][]{Tahani2022,Wu2024} have shown that in the plane of the sky, magnetic field lines lie predominately perpendicular to high column density ($\sim10^{22}$ cm$^{-2}$) filamentary structures which often are forming stars. To the best of our knowledge, we are the first to measure alignment of stellar kinematics and magnetic field lines. We hypothesize that in Perseus, the initial collapse of the cloud formed a sheet. During the sheet formation, magnetic field lines are dragged along the collapsing such that they are preferentially perpendicular to the sheet. As the sheets get denser, further collapse within this sheet into filamentary structures occurs perpendicular to the field lines. As these filaments begin forming stars, the kinematics the stars inherit from the collapsing gas will be predominately perpendicular to the magnetic field, producing the trend we observe. In Taurus, it is possible that the magnetic field was not affected by the collapse of the cloud as much, or that this trend in orientations was once present and has since been obscured by other dynamical effects. It is also worth noting that these trends are in projection, and more prominent correlations in Taurus may be present along the line of sight. 

\cite{Tahani2022} find that the 3D structure of the magnetic field in Perseus is perpendicular to the star-forming clouds. They attribute this morphology to the Per-Tau shell ``bending" the magnetic field as it expands, wrapping the field around the clouds. This would also produce the perpendicular $\delta\theta$ distributions we see in Perseus, as the field lines would partially trace the edge of the shell as the stars follow the shell's expansion direction. However, if this mechanism is the only process affecting the field/cloud morphology, we expect to see the same behavior in Taurus, as both of these groups are on the Per-Tau shell and their stars' kinematics retain a memory of this shell's expansion (Figure~\ref{fig:Per_taur_vsfs}). They also note how the overall field morphology they find is an approximation which neglects the small-scale field variations that we somewhat probe here. Future measurements of magnetic field strengths and orientations at smaller scales and in other young groups \citep[e.g. the JCMT BISTRO Survey,][]{Ward-Thompson2017}, as well as MHD simulations of star formation and stellar kinematic evolution in a Milky Way-like galaxy, will deepen our understanding of the trends we observe. 

\section{Limitations and Future Work}
\label{sec:final_limits_future}

Star formation is a result of ISM turbulence, gravity, supernovae, and magnetic fields over a wide range of length scales. Our work demonstrates that the kinematics of young stars reflect the complicated interplay of all these physical processes. Analyzing the stellar kinematics can help us better understand both the star formation process itself and the kinematics of the ISM that the stars are born out of. Our results show that the motions of young stars in many groups show characteristics of turbulence. Some show clear evidence of (an)isotropic expansion and/or recent supernova explosions. Remarkably, we have also identified a correlation between the motions of young stars and the direction of the magnetic fields in Perseus. 

Our work is limited by how the degeneracy of some of the physical processes (e.g., radial motion from gravity vs supernovae) manifest in the stellar kinematics. We are also limited by the measurement uncertainties and our sampling biases. We identify several potential biases and limitations of our analysis that restrict our ability to make definitive claims, and provide avenues of future work that may rectify these issues.

Gravitational interactions between these young stars can erase the memory of ISM turbulence, cloud collapse, SN explosions, or magnetic field effects through dynamical relaxation. This is most clearly seen in the VSFs of ONC and Orion A, as they are extremely flat up to the highest scales. Even though the dynamical times of each of our groups are in the tens to hundreds of Myr, compact sub-clusters within each group are likely flattening the VSFs at small scales ($\ell \lesssim 10$ pc). Our sample is biased toward these dense sub-clusters, as previously discussed in Section~\ref{sec:resTurb}. Removing poorly measured stars and stars with low membership probability preferentially removes stars in the more diffuse regions of these groups that may indeed have formed from the same cloud as the larger structures. 

However, our inclusion of cuts based on membership probability, radial velocity error, and parallax removes contamination from our samples. The use of Gaia DR3 to expand the number of stars per group when compared with \cite{Ha2021,Ha2022} in tandem with these cuts means that our sample is more complete and has lower noise. Further data releases and future telescopes/missions with higher measurement precision will allow this kind of analysis to include even fainter stars, which are currently below the detection limit of Gaia ($G \sim 21$) and/or APOGEE ($H \sim 12-13$) as well as poorly measured sources and stars at the outskirts of these groups. 

A more extensive stellar catalog will also improve our statistics in the measurement of magnetic field - stellar kinematics alignment, possibly enabling a more comprehensive study across different scales. In addition, high-resolution magnetic field maps in more star forming regions will allow us to probe the connections between the field and the stellar kinematics at smaller scales and in diverse environments. 

The heightened scatter in line of sight velocity measurements, which roughly lies along the x direction in the groups analyzed here, is likely enhancing the amplitudes of primarily the x direction 4D VSFs we observe in Figure \ref{fig:VSF}. Despite our attempt to remove this bias by scaling the errors in proper motions to the width of the line of sight velocity error distribution, we can't determine if the heightened VSFs are a result of a physical phenomenon or the intrinsic spread of the velocity measurements along the line of sight. To test how our results are sensitive to the quality of radial velocity data, we split our sample for each group in half around the median radial velocity error value. The VSFs of the noisier data tend to be flatter, as expected, but otherwise the main trends do not change. The main trends in the radial profiles and the magnetic field alignment are also insensitive to the split.

Without past works presenting compelling evidence for enhanced motion in this direction \citep[i.e., Orion C and D][]{Kounkel2018}, we cannot make claims about the amplitudes of the 4D VSFs. More precise radial velocity measurements are necessary to remove this systematic enhancement if 6D kinematic analyses of Milky Way stars are to be effective and bias-free. In tandem with improved observational data, analyzing high-resolution simulations of Milky Way-like galaxies and individual star-forming molecular clouds can be extremely helpful to untangle the complicated contributors to the kinematics of the young stars. 

Nonetheless, our VSF analysis shows that the anisotropy (difference along x, y, and z directions) of turbulence reflected in the motions of young stars is overall mild (even if it is not due to observational bias). This is generally consistent with the gas kinematics analysis of the ISM \citep[e.g.,][]{Larson1981,ChepurnovnLazarian2010}. Moreover, \cite{Zhou2022} used a catalog of Class I and II YSOs to probe the kinematics of their natal clouds. They found that they display isotropic motions, despite anisotropic density structures, using velocity dispersion to reproduce Larson's Relation in two directions as opposed to the 6D and 4D VSFs we use here. On the other hand, our radial profile analysis shows that in several groups (e.g., Orion BCD, $\lambda$ Ori, and Upper Sco), the motions of the young stars are clearly anisotropic. 

To further disentangle the different physical processes affecting stellar kinematics, we plan to extend our analysis on the same groups with stars binned by age and include a sample of YSOs \citep[e.g.,][]{Marton2016} associated with these star forming regions. This will help observationally separate the many overlapping phenomena affecting gas and stellar kinematics that produce the anisotropy. Including a sample of YSOs will also allow us to probe tracers of ISM kinematics that have not had many strong dynamical interactions with one another. 

\section{Conclusions}
\label{sec:conclusion}
We present a multi-faceted investigation into several star forming regions in the Milky Way, analyzing the contributions of turbulence, gravity, supernovae, and the magnetic field to the kinematics of young stars. Using Gaia DR3 astrometry in tandem with APOGEE DR17 radial velocity measurements, we subdivide our stellar sample into eight associations: Orion (split into Orion BCD, ONC, Orion A, and $\lambda$ Ori), Upper Sco, Taurus, and Perseus. We calculated the 6D and 4D VSFs of each group to understand the internal velocity statistics. We looked for signatures of turbulence, identified signs of local energy injection, and uncovered evidence of anisotropic motions. We also computed radial (from the center of mass of each group) profiles for these groups to compliment the VSFs and searched for anisotropic expansion trends from either previous gravitational collapse/interactions or supernovae. Finally, we compared the magnetic field orientations to stellar proper motions in Taurus and Perseus for correlations in their directions in the plane of the sky. The results of our work are summarized as follows:

\begin{itemize}
    \item We see mild anisotropy in the VSFs of a few ($\lambda$ Ori, Taurus) groups we analyze, but on the whole the VSFs seem to imply isotropic motion with a variety of slopes ranging from $\sim0$ to $\sim1$ at various scales. Several groups (Orion, $\lambda$ Ori, Upper Sco, Taurus, Taurus + Perseus) show peaks in their VSFs, indicative of local energy injection. 
    \item There are clear anisotropic expansion trends in the radial profiles of several (Orion BCD, $\lambda$ Ori, Taurus) groups where the expansion is only observed in certain directions and not in others. In the groups that show mild anisotropy in their 4D VSFs, we see the most dramatically anisotropic expansion trends. 
    \item We find a correlation between the magnetic field and stellar proper motions in Perseus. The stars are moving preferentially perpendicular to the local magnetic field. 
\end{itemize}

We provide several interpretations for what could be causing these trends:

\begin{itemize}
    \item Broadly, the groups we analyze display stellar kinematics that are consistent with Larson's Relation at intermediate length scales ($\ell \sim 10-100$ pc). However, gravitational interactions across scales also encode a variety of slopes in the VSFs, with relaxed groups showing a slope of 0 (ONC and Orion A) and expansion trends from crossing after group-wide collapsing or simply stellar drifting manifest as a slope closer to 1 ($\lambda$ Ori and Taurus). The non-spherical distribution of the natal gas and resulting stars in their respective groups makes larger scale gravitational collapse happen asymmetrically, potentially leading to the anisotropy we observe in Taurus. Sub-group motions and collisions could also create anisotropic kinematics, as is likely the case in Orion BCD, $\lambda$ Ori and Perseus. 
    \item Energy injected from a single supernova is observed in the velocity structure functions of Orion BCD and $\lambda$ Ori, and multiple generations of supernovae and their effect on stellar kinematics are observed in the combined VSFs of Taurus and Perseus. The non-uniform distribution of stars along the shells produce some of the observed anisotropic motions and expansions. Our estimates of the ages of these SN are consistent with the literature. 
    \item We hypothesize that the correlation between the magnetic field direction and stellar kinematics in Perseus is a result of the initial cloud collapse altering the magnetic field morphology such that the field lines are preferentially perpendicular to filaments (and therefore the stellar kinematics). More work is required to understand the deeper physical connections between magnetic fields and the young stellar kinematics. 
\end{itemize}

We want to emphasize that in each of the groups we analyze we find  multiple, overlapping causes for the observed kinematic trends. The physical phenomena that create the trends we see have been known to be driving the dynamics of the ISM and affecting the star formation process. It is clear that some of these effects are indeed causing anisotropic motion in these groups. Our analysis demonstrates that these young stars retain strong memories of the conditions before their formation that we can probe using their kinematics. 

\appendix
\begin{figure*}[bht]
    \centering    
    \includegraphics[width=0.77\linewidth]{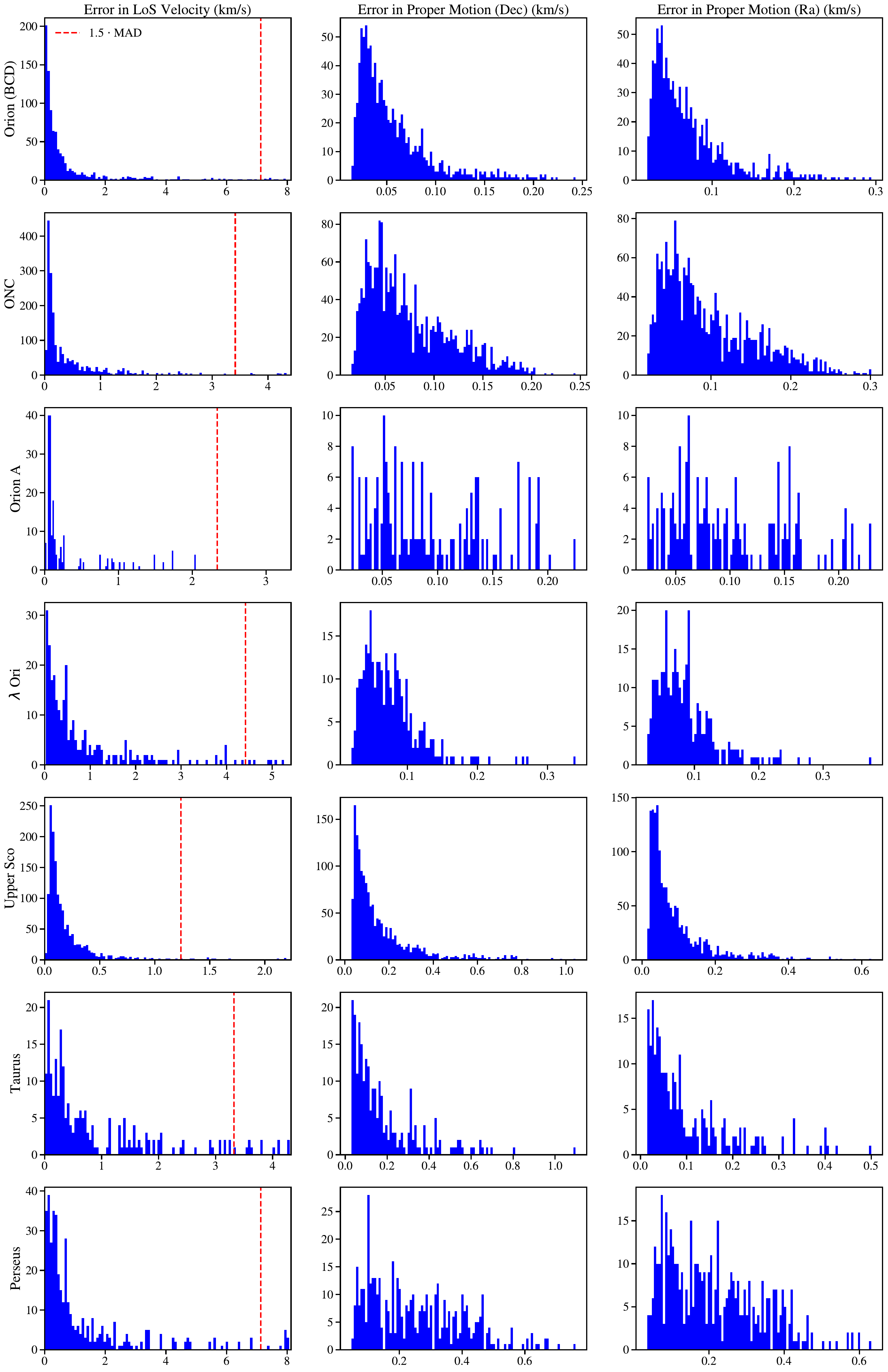}
    \caption{The velocity error distributions in different directionsF5 (left: radial velocity, center: proper motion in declination, right: proper motion in right ascension) for each group analyzed in this work. The red dashed line in the leftmost column (radial velocity error) is where we cut our sample to remove data with high radial velocity errors. }
    \label{fig:velo_err_dists}
\end{figure*}

\begin{acknowledgements}

We thank the anonymous referee for their valuable suggestions. We would like to thank Mark Krumholz, Mark Heyer, Mateusz Ruszkowski, and Shmuel Baily for helpful discussions. Y.L. acknowledges financial support from NSF grants AST-2107735 and AST-2219686, NASA grant 80NSSC22K0668, and Chandra X-ray Observatory grant TM3-24005X. HL is supported by the National Key R\&D Program of China No. 2023YFB3002502, the National Natural Science Foundation of China under No. 12373006, and the China Manned Space Program with grant No. CMS-CSST-2025-A10.

\end{acknowledgements}

\software{ \texttt{NumPy} \citep{numpy2011,numpy2020}, \texttt{Matplotlib} \citep{matplotlib}, \texttt{Astropy} \citep{astropy}, \texttt{emcee} \citep{emcee}}

\bibliographystyle{aasjournal}
\bibliography{refs}
\end{document}